\documentclass[aps,superscriptaddress,amsmath,amssymb,floatfix,twocolumn,showpacs,amsfonts,longbibliography]{revtex4-2}

\usepackage{graphicx}
\usepackage{dcolumn}
\usepackage{bm}
\usepackage{physics}
\usepackage{times}
\usepackage[varg]{txfonts}
\usepackage{textcomp}
\usepackage{graphicx}
\usepackage{subfigure}
\usepackage{tabu}
\usepackage{color}
\usepackage[colorlinks=true,citecolor=blue,urlcolor=blue,linkcolor=blue,hyperindex]{hyperref}
\usepackage{braket}
\usepackage{overpic}
\usepackage{amssymb}
\usepackage{multirow}
\usepackage{verbatim}
\title{My Title Here}
\usepackage[]{placeins}
\usepackage{ulem}

\usepackage[usenames,dvipsnames]{xcolor}
\usepackage{tcolorbox}
\usepackage{tabularx}
\usepackage{array}
\usepackage{colortbl}
\tcbuselibrary{skins}
\usepackage{multirow}

\newcommand{\dt}[1]{\frac{\partial #1}{\partial T}}

\newcommand{\dbdt}[0]{\frac{E(\textbf{k})\beta }{k_B T^2}}

\newcolumntype{Y}{>{\raggedleft\arraybackslash}X}

\tcbset{tab1/.style={fonttitle=\bfseries\large,fontupper=\normalsize\sffamily,
colback=yellow!10!white,colframe=red!75!black,colbacktitle=Salmon!40!white,
coltitle=black,center title,freelance,frame code={
\foreach \n in {north east,north west,south east,south west}
{\path [fill=red!75!black] (interior.\n) circle (3mm); };},}}

\tcbset{tab2/.style={enhanced,fonttitle=\bfseries,fontupper=\normalsize\sffamily,
colback=white,colframe=blue!0!black,colbacktitle=Salmon!40!white,
coltitle=black,center title}}

\begin{document}

\preprint{APS/123-QED}
\title{Topological properties of multilayer magnon insulators}
\author{Stephen Hofer}
\affiliation{Physics Department, 1245 Lincoln Drive, Southern Illinois University, Carbondale, IL 62901, USA}

\author{Trinanjan Datta}
\email[Corresponding author:]{tdatta@augusta.edu}
\affiliation{Department of Chemistry and Physics, Augusta University, 1120 15$^{th}$ Street, Augusta, Georgia 30912, USA}

\author{Sumanta Tewari}
\affiliation{Department of Physics and Astronomy, Clemson University, Clemson, South Carolina 29634, USA}

\author{Dipanjan Mazumdar}
\email[Corresponding author:]{dmazumdar@siu.edu}
\affiliation{Physics Department, 1245 Lincoln Drive, Southern Illinois University, Carbondale, IL 62901, USA}


\date{\today}
\begin{abstract}
Two-dimensional magnetic insulators can be promising hosts for topological magnons. In this study, we show that ABC-stacked honeycomb lattice multilayers with alternating Dzyaloshinskii-Moriya interaction (DMI) reveal a rich topological magnon phase diagram. Based on our bandstructure and Berry curvature calculations, we demonstrate jumps in the thermal Hall behavior that corroborate with topological phase transitions triggered by adjusting the DMI and interlayer coupling. We connect the phase diagram of generic multilayers to a bilayer and a trilayer system. We find an even-odd effect amongst the multilayers where the even layers show no jump in thermal Hall conductivity, but the odd layers do. We also observe the presence of topological proximity effect in our trilayer. Our results  offer new schemes to manipulate Chern numbers and their measurable effects in topological magnonic systems.
\end{abstract}

\maketitle


\section{\label{sec:secI}Introduction}
{\flushleft T}he discovery of two-dimensional magnetic crystals in the past few years \cite{Huang2017,Gong2017,Deng2018,Mak2019,Gibertini2019, McGuire2017} has raised the prospect of realizing topologically protected magnons (spin-wave excitations)~\cite{Loss2017,Wang2018}. Since topological materials exhibit robustness against disorder~\cite{HasanKane,Ando,Moore2010}, compared to their electronic counterpart, the existence of topologically protected magnonic edge states can potentially lead to the realization of much lower power consumption spintronic devices~\cite{Andreas2018,Chumak2015,Bauer2012,Uchida2010,Cornelissen2015,Wang2019,Avci2017,Kovalev,MookEdge,Chisnell_Kagome} and applications in quantum information science~\cite{Paolo2017}. Recently, it has been theoretically predicted~\cite{OwerreMono,Lifa2013,MookEdge} and experimentally demonstrated~\cite{Onose297, Chisnell_Kagome} that it is possible to harbor topological magnon edge states in realistic geometrically frustrated magnets. At present various materials have the potential to host topological magnonic states~\cite{Chisnell_Kagome,ChenDai,Miura,Tsirlin,Lifa2013,MookEdge}, including the honeycomb magnetic halide CrI$_3$~\cite{ChenDai}, spin-1/2 Heisenberg antiferromagnets Na$_3$Cu$_2$SbO$_6$ ~\cite{Miura} and $\beta$-Cu$_2$V$_2$O$_7$~\cite{Tsirlin}. In addition to the honeycomb lattice, topological magnon excitations have been proposed to exist in the kagom\'{e} magnet system Cu (1-3, bdc)~\cite{ChisnellPRB} and the square lattice geometry~\cite{PhysRevB.99.054422}. Topological phase transition induced by magnetic proximity effect in CrI$_3$/SnI$_3$/CrI$_3$ trilayer has been proposed~\cite{Zeng_2019}. Einstein-de Haas effect of topological magnons has also been predicted~\cite{edH}.

A topological magnon insulator (TMI) is the bosonic analog of the quantum spin Hall state~\cite{Haldane,PhysRevB.93.134502,PhysRevLett.117.227201}. This phase is fundamentally different from topological magnetic insulators wherein topological electronic insulators are doped with magnetic 3$d$ atoms~\cite{Tokura2019}. The topological origins of the bosonic TMI phase can be traced to spin-orbit coupling interaction which manifests itself in the form of Dzyaloshinskii-Moriya interaction (DMI)~\cite{Onose297} and/or pseudodipolar interaction~\cite{wangrev,Wangpyro,Wanghc,Wangdmi}. 
Typically, the later interaction occurs in compounds with heavy ions, a class of material which is beyond the scope of our current investigation~\cite{Wang2018}.

The experimental realization of monolayer, bilayer and few-layer CrI$_3$ with tunable magnetic properties \cite{Huang2017,Jiang2018,Thiel2019,Li2019,Song2019,Li2019} provides materials science motivation to pursue a study of few-layer coupled bosonic topological magnon system. It has been shown that protected magnon states in the AB-stacked bilayer honeycomb propagate in the same (opposite) direction for ferromagnetically (antiferromagnetically) coupled layers~\cite{OwerreAB}. Furthermore, Andreas~$et~al.$~\cite{Andreas2018} demonstrated through numerical calculations that the edge currents are robust again weak disorder compared to the bulk current in normal metal/TMI/normal metal heterostructure. 
 
\begin{figure*}[hbtp]
\centering
      \includegraphics[width=17.8cm]{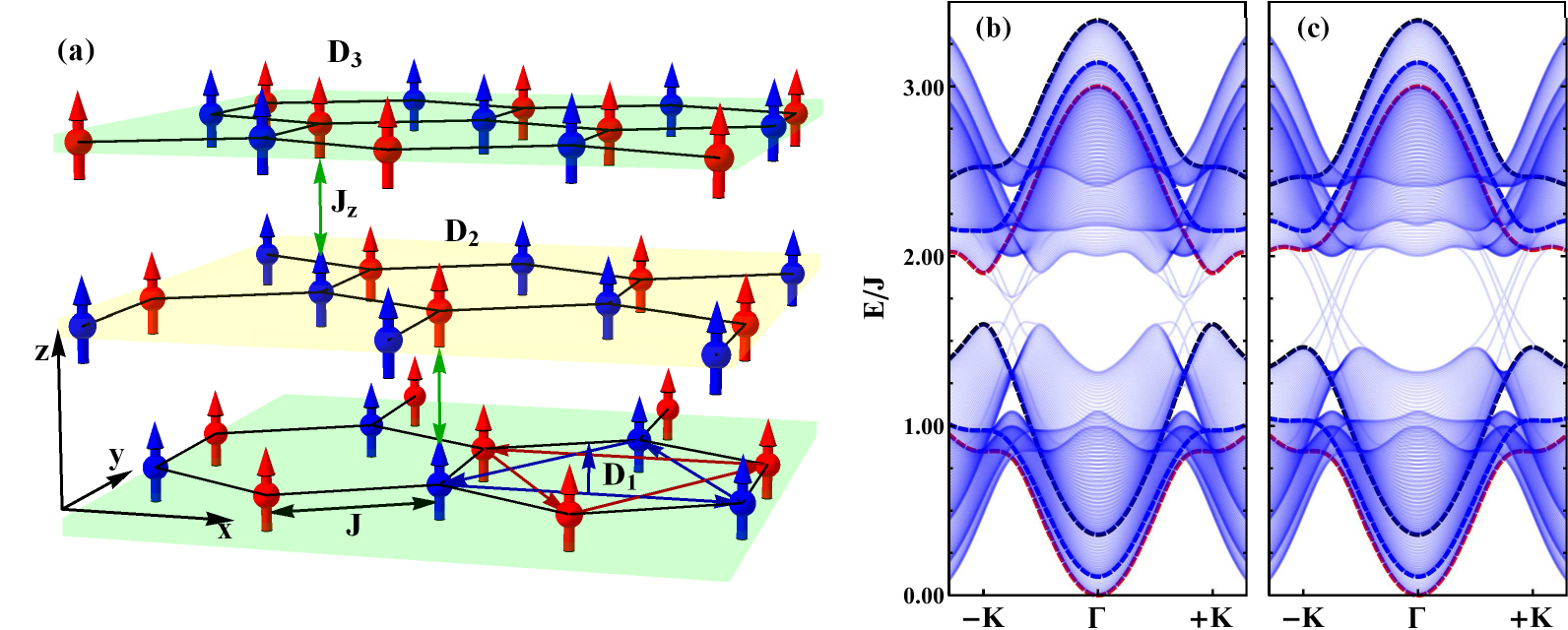}
      \caption{Trilayer configuration, bandstructure, and edge states.~\textbf{a} Lattice crystal structure with ferromagnetic spin ordering. Spin sites A(B) are denoted with red(blue) spheres. $J$ denotes intralayer nearest-neighbor ferromagnetic Heisenberg exchange interaction. $D_\tau$ denotes the layer specific next-nearest neighbor DMI, where $\tau \in (1,2,3)$ indexes the layer. Although we show distinct DMI interactions for each layer, for our calculations we will take $D_1 = D_3$ (reason explained in main text). $J_z$ denotes the interlayer Heisenberg exchange interaction. ~\textbf{b}--\textbf{c} Bulk bands (solid dashed lines) with edge states (thin blue lines) for the trilayer plotted along $k_y=0$. The parameters are $J=2J_z=4D_1=4D_3$, $D_2/D_1=-0.426$, and $D_2/D_1=0.34$, respectively}.
      \label{fig:crystal01}
\end{figure*}

We investigate the thermal transport properties of  ferromagnetically coupled TMI multilayers with different DMI strength in adjacent layers, as shown in Fig. \ref{fig:crystal01}. Such topologically distinct layers lead to the possibility of observing several TMI phases. The presence of DMI interaction in a magnetic system without inversion center will create band gaps in the magnon dispersion relation~\cite{MookEdge} and impart non-trivial topological nature to the system. The topological texture of these bands give rise to a non-vanishing Berry curvature. The physical consequence is a nonzero topological invariant (Chern number and winding number) that directly influences thermal Hall conductivity~\cite{MookEdge,MookHall}. The emergence of TMI phases are characterized by jumps in the thermal Hall conductance that are analogous to the electrical Hall conductance jumps in Quantum Hall systems. 

Using spin wave theory we compute the topological band structure and its edge states, Chern number, and transverse thermal hall conductance $\kappa_{xy}$ behavior. We show that the multilayer supports a rich phase diagram which can be explored by tuning the strength of the intermediate layer's DMI ($D_2$ in Fig.~\ref{fig:crystal01}) relative to the top and the bottom layers or by adjusting the interlayer interaction strength $J_z$. We investigate and discuss the variation in thermal Hall conductance with changing interlayer DMI strength ratio $D_2/D_1$ and for different interlayer coupling relative to the DMI interaction $J_z/D_1$. Furthermore, we show that the physics of few-layered topological multilayer has its own characteristic transport properties. The presence of an uncompensated topological layer in odd layered configuration leads to non-trivial behavior in the thermal Hall conductance behavior. As a result, we show that there is an odd-even layering effect which manifests itself as a jump or not in the transverse thermal Hall conductance behavior. Additionally, the trilayer exhibits a topological proximity effect which can be induced by external pressure. Overall, we put forward the design and characterization of a finite number of layered topological magnon insulator systems (odd or even) with several interesting effects directly related the the topology of the system. 

This article is organized as follows. In Sec.~\ref{sec:secII} we present the model and the method. In Sec.~\ref{sec:secIII} we perform the topological characterization of our multilayer. In Sec.~\ref{sec:secIV} we present our thermal Hall response results of the multilayer system. Finally, in Sec.~\ref{sec:secVI} we present our conclusions.

\section{\label{sec:secII} Model and Method}

{\flushleft W}e analyze an $ABC$-stacked multilayer honeycomb lattice which is consistent with the low temperature (rhombohedral) experimental structure of CrI$_3$~\cite{McGuire2017,Mijin2018}. To connect with conventional experimental sandwich structures, the DMI strength alternates between two values (for example, {$D_1=D_3$ and $D_2$} in Fig.~\ref{fig:crystal01}).The individual layers are chosen to be ferromagnetically aligned which is consistent with bulk and odd layered CrI$_3$~\cite{Huang2017,Thiel2019}. While experimental evidence suggests that even layered CrI$_3$ shows a net antiferromagnetic configuration in the ground state~\cite{Thiel2019}, it has also been demonstrated that transition to the ferromagnetic state can be induced using external pressure~\cite{Li2019,Song2019}. Later, we will show that the trilayer forms the basic building block for all odd layered configurations (five, seven, etc) while the bilayer is the basic building block of all even layered structures (four, six, etc).  

We model our two dimensional multi-layer FM topological insulator using the Heisenberg exchange term $H_{FM}$ and the DMI term $H_{DMI}$. To model our few layer system we add an interlayer interaction $H_{int}$ to stack the monolayers, as seen in Fig.~\ref{fig:crystal01}a.  The generic multilayer Hamiltonian can be written as
\begin{equation}
  H=H_{FM}+H_{DMI}+H_{int},
\end{equation}
where the individual terms are given by the following expressions
\begin{subequations}
\label{eq:ham}
\begin{align}
& H_{FM}=-\sum_{<\alpha,\beta>}\sum_{\tau=1}^L J_{\tau} \bf{S}_{\tau,\alpha} \cdot\bf{S}_{\tau,\beta}, \\
& H_{DMI}=\sum_{<<\alpha,\beta>>}\sum_{\tau=1}^L D_{\tau} \bf{\hat{z}}\cdot\left(\bf{S}_{\tau,\alpha} \times \bf{S}_{\tau,\beta} \right),\\
& H_{int}=-\sum_{\alpha,\beta}\sum_{\tau=1}^{L-1} J^{\tau,\tau+1}_{\alpha,\beta}\bf{S}_{\tau+1,\alpha} \cdot \bf{S}_{\tau,\beta}.
\end{align}
\end{subequations} In the above equations $\tau$ indexes the layer, $\alpha$ and $\beta$ index the sublattice degrees of freedom, $J_\tau$ is the intralayer ferromagnetic exchange, ${\bf S}_{\tau,\alpha}$ is the site-specific spin moment , $D_\tau$ is the next-nearest neighbor DMI, and $J^{\tau,\tau+1}_{\alpha,\beta}$ is the ferromagnetic interlayer exchange.  In our $ABC$-stacked trilayer honeycomb lattice $\verb+{+\alpha,\beta\verb+}+\in\verb+{+A,B\verb+}+$, $\tau\in\verb+{+1,2,3\verb+}+$, $J_\tau=J_{\tau^\prime}\equiv J$, $D_1=D_3\neq D_2$, and $J^{\tau,\tau+1}_{B,A}=J_z$ with all other $J^{\tau,\tau+1}_{\alpha,\beta}=0$. The interlayer interaction depends on the stacking arrangement. Our choice of magnetic interaction (exchange and DMI) parameters are guided either by CrI$_3$~\cite{ChenDai} system or is based on the choice of physically reasonable model parameters. While magnetic anisotropy plays an important role in the magnetic ordering of 2D magnets such as CrI$_3$, its contribution to the magnonic bandstructure serves to raise or lower the energy of each band by an equal amount. No new band crossings are observable as a result of this interaction, so it is omitted to simplify the model. 

Next, we apply linear spin wave theory transformation to Eqs.~\eqref{eq:ham} and Fourier transform the Hamiltonian. Thus, the momentum space Hamiltonian can be written as $H=$ $\sum_\textbf{k}\Psi^\dagger_\textbf{{k}}\mathcal{H}(\textbf{k})\Psi_\textbf{k}$, with the basis vector $\Psi^\dagger_\textbf{k}=\left(b^\dagger_{A,1,\textbf{k}},b^\dagger_{B,1,\textbf{k}}, \cdots, b^\dagger_{A,L,\textbf{k}},b^\dagger_{B,L,\textbf{k}}\right)$. Specifically, for our trilayer configuration the Hamiltonian takes the form \begin{equation}
    \mathcal{H}(\textbf{k})=
    \begin{pmatrix}\mathcal{A}_1(\textbf{k}) & \mathcal{B}(\textbf{k}) & 0 \\
    \mathcal{B}^\dagger(\textbf{k}) & \mathcal{A}_2(\textbf{k}) & \mathcal{B}(\textbf{k})\\
    0 &  \mathcal{B}^\dagger(\textbf{k}) & \mathcal{A}_3(\textbf{k})
    \end{pmatrix},
\end{equation} where $\mathcal{A}_i(\textbf{k})$ and $\mathcal{B}(\textbf{k})$ are 2 $\times$ 2 matrices that describe the intralayer and interlayer interactions, respectively. Note, for a $L$-layered system the Hamiltonian matrix would be $2L \times 2L$ in dimension. The intralayer interaction  $\mathcal{A}_i(\textbf{k})$ is given by
\begin{equation}
\label{eq:AMat}
     \mathcal{A}_{\tau}(\textbf{k})=\begin{pmatrix}\Theta^\tau_A+D_\tau S m(\textbf{k}) & -JS f(\textbf{k})\\
    -JS f^*(\textbf{k}) &\Theta^\tau_B -D_\tau S m(\textbf{k})
    \end{pmatrix},
 \end{equation} where $\Theta^\tau_\alpha = 3JS+ \theta^\tau_\alpha J_zS$, implying $\theta^1_A=\theta^3_B=0$ and  $\theta^1_B=\theta^2_A=1=\theta^2_B=1=\theta^3_A=1$. The explicit interlayer coupling expression is given by
\begin{equation}
\label{eq:BMat}
     \mathcal{B}(\textbf{k})=\begin{pmatrix}-J_{AA}S f^*(\textbf{k}) & -J_{AB}S f(\textbf{k})\\
    -J_{BA}S & -J_{BB}S f^*(\textbf{k})
    \end{pmatrix}=\begin{pmatrix}0 & 0\\
    -J_zS & 0
    \end{pmatrix},
 \end{equation} where $f(\textbf{k})$=$\sum_ie^{-i\textbf{k}\cdot \vec{\delta}_i}$ is the nearest neighbor structure factor. The lattice position vectors $\vec{\delta}_i$ are given by $\vec{\delta}_i\in \{(0,-1),(\sqrt{3}/2,1/2),(-\sqrt{3}/2,1/2)\}$. The anti-symmetric next-nearest neighbor structure factor corresponding to the DMI term is given by $m(\textbf{k})$=$\sum\limits_{i} 2\sin(\textbf{k}\cdot\mathbf{\rho}_{i})$ where $\vec{\rho}_i\in \{(\sqrt{3},0),(-\sqrt{3}/2,3/2),(-\sqrt{3}/2,-3/2)\}$. 
 
The trilayer bulk and edge configuration bandstructure is shown in Figs.~\ref{fig:crystal01}b and \ref{fig:crystal01}c. The TMI bandstructure with edge states has differences from its electronic counterpart. Inspecting Figs.~\ref{fig:crystal01}b and \ref{fig:crystal01}c we observe some interesting differences between our bosonic TMI and an electronic or magnetic-TI. While the gap and edge states are approximately around zero energy for fermionic systems, in the bosonic case the gap is located at a higher energy. Furthermore, from the nature of the edge states we get a hint that the two panels belong to different topological phases. In fact, under appropriate external tuning the trilayer can undergo a topological phase transition (TPT) from panel (b) to (c). To track these TPTs we employed a methodical approach of searching for band gap closings. We computed gap closings specifically at the high symmetry $\pm K =(\pm 4\pi/3\sqrt{3},0)$ in the Brillouin zone. At this momentum point, the nearest-neighbor structure factor $f(\textbf{k})$ becomes zero. This eliminates the contribution of our strongest interaction $J$. Thus, the energy scale of the problem is governed by $D_1$, leaving $D_2$ and $J_z$ as the tuning parameters by which we can explore the various topological phases of our system.

We define a multilayer tuning ratio $\delta= D_2/D_1$. This will serve as a control knob to study TPTs. As we show later the $\delta=1$ configuration is of particular interest because of its feasibility to be naturally realized in an experimental setup. The Chern numbers are rearranged at a TPT. Since, the interband edge states are a consequence of these Chern numbers, a change in them implies that the number of edge states will alter across a transition. This is clearly visible in Figs.~\ref{fig:crystal01}b and ~\ref{fig:crystal01}c. For example, the number of interband edge states in Fig.~\ref{fig:crystal01}b is one, while in Fig.~\ref{fig:crystal01}c the number is three. The main physical property that emerges from the TMI phase is the existence of these chiral magnonic edge states which contribute to the non-vanishing thermal Hall conductivity~\cite{Onose297,MookHall,Kovalev_2017_Hall,OwerreAB,MurakamiPRB,MurakamiPRL}. In the next section, we will study the nature of these TPTs in more detail.

 \section{\label{sec:secIII}Topological Characterization}
\begin{figure*}[t]
      \centering
      \includegraphics[width=17.8cm]{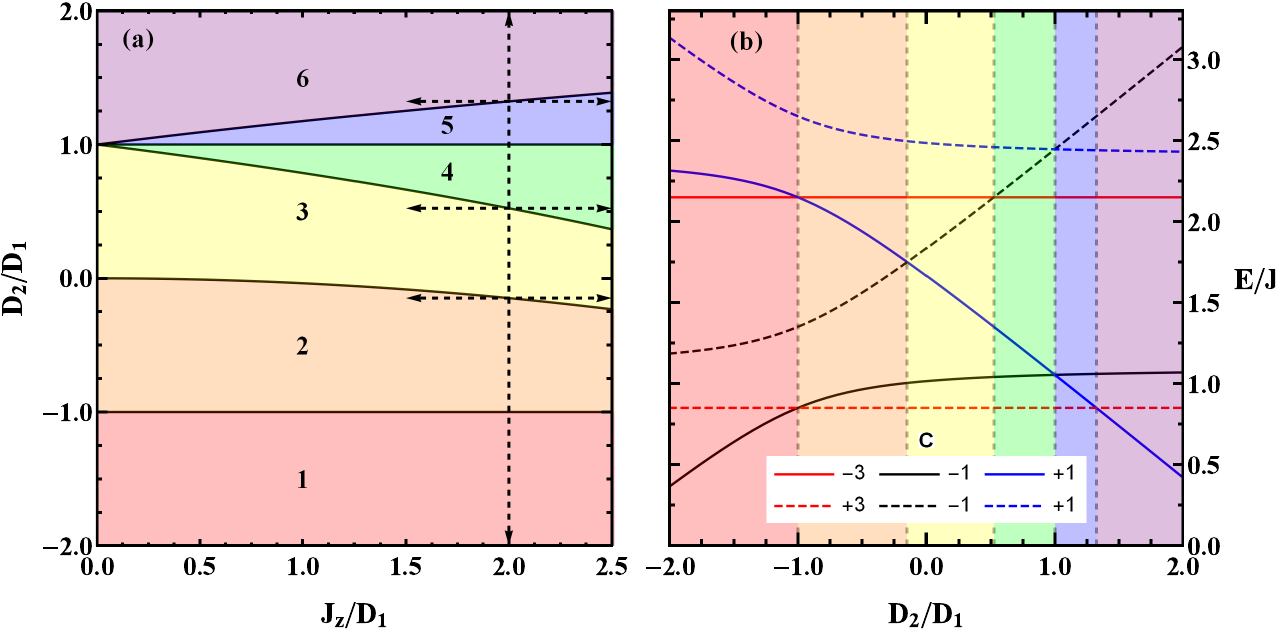}
      \caption{Topological phase diagram of a trilayer.~\textbf{a} Each phase is separated by gap closings corresponding to the $\delta_n$ represented by the solid black lines. In this parameter range there are six distinct phases shown. Dashed lines represent directions in which the thermal Hall effect is analyzed. \textbf{b} Energy eigenvalues of the system at $\pm K$ as a function of $D_2/D_1$. The ratio  $J_z/D_1=2$, corresponds to the vertical dashed line in panel (a). The Chern number for each band associated with the eigenvalue is indicated in the legend.}
      \label{fig:TPT}
  \end{figure*}

  {\flushleft T}he trilayer topological phase diagram is shown in Fig.~\ref{fig:TPT}.  For convenience, the phases are color coded so that we can compare the two panels (a) and (b). The result depends on two interaction ratios, one is $\delta$ and the other $J_z/D_1$. The feasibility of tuning $J_z$ using pressure has already been experimentally demonstrated in a hexagonal lattice system~\cite{Li2019,Song2019}. Based on our studies, we show that there can be further motivation to tune the DMI interactions, too. For suitable parameter ranges we observe a quantum Hall behavior in our proposed bosonic system. 
To track the TPTs we compute the energy eigenvalues at the $\pm K$ high symmetry points. The analytical expression for the energy eigenvalues calculated at $+K$ yields
\begin{equation}
E^{(3)}_\eta = 
     \begin{cases}
    \frac{3J}{2}+(-1)^{\eta+1}\frac{3\sqrt{3}}{2}D_1 &\eta = 1,2 \\
       \frac{3J}{2}+\frac{J_z}{2}+(-1)^{\eta+1}\Delta+\frac{3\sqrt{3}}{4}D_1(1-\delta) & \eta = 3,4\\
    \frac{3J}{2}+\frac{J_z}{2}+(-1)^{\eta+1}\Delta-\frac{3\sqrt{3}}{4}D_1(1-\delta)   & \eta = 5,6 
    \end{cases}
     \end{equation} where $E^{(3)}_{\eta}$ are the trilayer eigenvalues and we have defined $2\Delta = \sqrt{J^2_z+27D_1^{2}\left(1+\delta\right)^2}$. Eigenvalues for the system solved at $-K$ result in the same solutions as above, except with a sign change which relabels $\eta =3,4$ to $\eta = 5,6$ and vice versa. Just for comparison purposes, we list the energy solutions for the bilayer problem in Appendix \ref{sec:appendixA}, see Table \ref{tab:EigL2}. Next, the TPTs are obtained from the real solutions of $E^{(3)}_i=E^{(3)}_j$ with $i\neq j$ using the above expressions. The topological phase boundaries can be defined as
\begin{equation}
\delta_n = 
     \begin{cases}
       \quad \frac{6\sqrt{3}+3\frac{J_z}{D_1}}{6\sqrt{3}+\frac{J_z}{D_1}},
       &\quad n = 5 \quad \text{where} \quad \delta_5\in (1,3) \\
       \quad 1, &\quad n =4 \\
       \quad \frac{6\sqrt{3}-3\frac{J_z}{D_1}}{6\sqrt{3}-\frac{J_z}{D_1}},
       &\quad n =3 \quad \text{where} \quad \delta_3\notin (1,3)\\
    \quad \frac{-J_z^2}{27D_1^2},&\quad n =2 \quad \text{where} \quad \delta_2\in (-\infty,0)\\
    \quad -1. &\quad n =1
     \end{cases}
     \end{equation}
 The number $n$ signifies the boundaries of the different phases. In the limit of zero interlayer interaction we can set $J_z=0$. In this case there are three phase boundaries separated by $\delta_n = -1, 0$, and $1$.

In Fig.~\ref{fig:TPT} we plot the six different phases based on the above solutions. The phase diagram depends on the ratio of $D_2/D_1$ (which can be positive or negative) versus $J_z/D_1$ variation. When $\delta$ is positive the DM interactions are aligned in the same direction. In this regime of the tuning parameters we find four phases (marked as 3 ---6 in the phase diagram). Whereas, when $\delta$ is negative, there are three phases (marked as 1 -- 3 in the phase diagram). Furthermore, around the $\delta =0$ line (FM middle layer) an interesting behaviour happens. This phase boundary between 2 and 3 varies as $J^{2}_{z}/D^{2}_{1}$. Hence, When $J_z < D_1$ (weak) the middle layer retains its non-topological behavior because the phase boundary mildly deviate from the $\delta=0$ line. However, for $J_z > D_1$ (strong) 
the $D_2$ deviates from zero to acquire a non-zero value. Thus, the FM layer starts to obtain a topological nature. We interpret this to be a signature of topological proximity effect displayed by the multilayer which can be experimentally realized by applying pressure~\cite{Li2019,Song2019}. For positive $\delta$ and for high $J_z > D_1$ we find that there are multiple phases into which the trilayer can transition into. These phases can be classified based on Chern numbers as we describe next, which are calculated from the Berry curvature in the following ways. For the Berry curvature calculation, we employ the following equation
 \begin{equation}\label{eq:BC}
     \Omega^{xy}_n(\textbf{k}) = -2\sum_{m\neq n}\Im \left[\frac{\bra{n}\frac{\partial \mathcal{H}(\textbf{k})}{\partial k_x}\ket{m}\bra{m}\frac{\partial \mathcal{H}(\textbf{k})}{\partial k_y}\ket{n}}{\big{[}E_n(\textbf{k})-E_m(\textbf{k})\big{]}^2}\right],
 \end{equation} obtained from standard perturbation theory approach \cite{Ando}. The Berry curvature calculation will be used later to compute the thermal Hall conductance. The Chern number is then calculated as
\begin{equation}\label{eq:Chern}
     C_n=\frac{1}{2\pi}\int_{BZ}\Omega_n^{xy}(\textbf{k})dk_xdk_y.
\end{equation}

In Fig.~\ref{fig:TPT}b we show the variation of the energy eigenvalues for $\delta$ at $J_z/D_1=2$ (shown as a dashed vertical line). This ratio choice is motivated by CrI$_3$ experimental parameters reported in Ref.~\cite{ChenDai}, where $J_z/D_1\approx 2$. We notice that the energy eigenvalues interchange indicating the presence of potential TPTs verified by the reordering of Chern numbers. The values for the Chern numbers given in Table~\ref{tab:ChernTab} can be generated by ordering the Chern numbers of each eigenvalue from the lowest to highest energy within each shaded phase. In our multilayer system there are contributions from several underlying bulk bands which can support topologically protected edge states. The Chern numbers determine the character of these edge states based on the winding number, defined as the partial sum $\nu_i=\sum_1^i C_i$. The winding number determines the number and chirality of the edge states which lie between the $i$th and $i+1$th bulk band. These states (as mentioned earlier) lie above the zero of energy.

If we adopt a fermionic classification scheme for the trilayer, then based on the winding number calculation,  $\nu_3=C_1+C_2+C_3$ we should have only two phases. The first two phases will have a winding number in the large gap between the lower and upper grouping of bands of $\nu_3=1$. The last four will have $\nu_3=3$ as documented in Table~\ref{tab:ChernTab}. The band crossing just below $\delta=0$ also accounts for the winding number $\nu_3$ change demonstrated by the number of topological edge states seen in the large gap between Figs.~\ref{fig:crystal01}b and \ref{fig:crystal01}c. However, we find that there are six distinct topological phases in Fig.~\ref{fig:TPT}a  with five transitions. So, in order to correctly identify all distinct topological bosonic phases we need to track the unique ordering of Chern numbers on either side of the topological phase boundary. We use this classification scheme to distinguish the different phases.

\begin{figure}[!t]
     \centering
     \includegraphics[width=8.6cm]{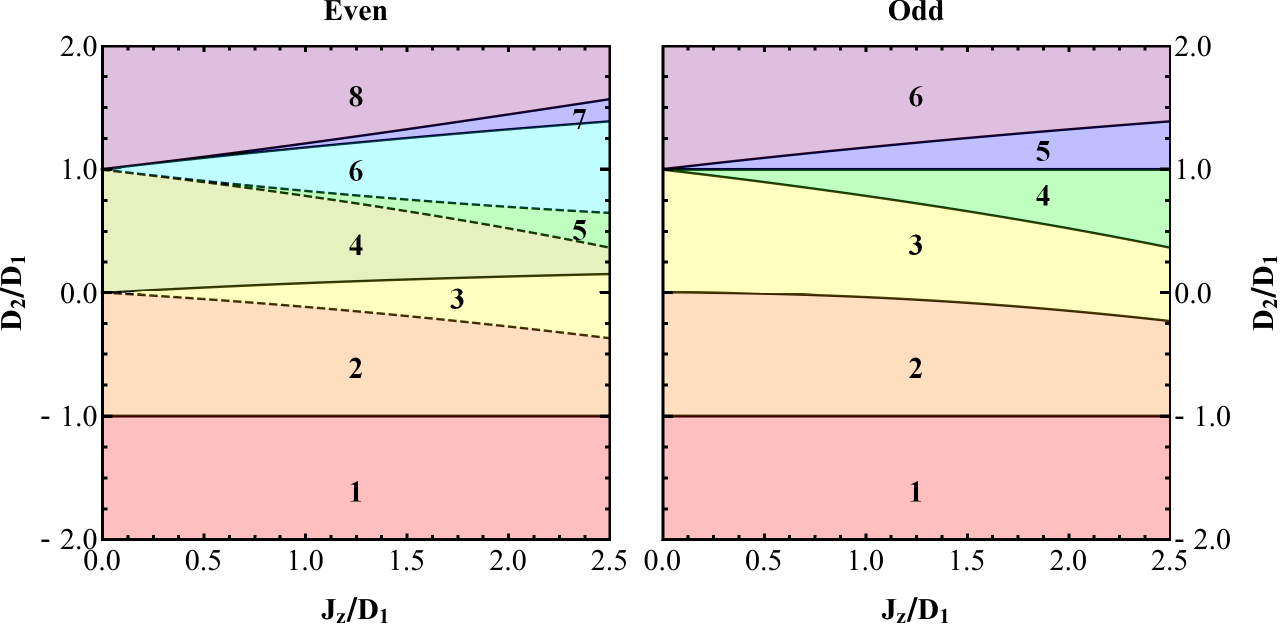}
     \caption{Phase diagram of the even and odd layered heterostructures. Solid (dashed) lines in the even layered phase diagram correspond to topological phase transitions associated with gap closings at +(-)K in the Brillouin zone.}
     \label{fig:eophase}
 \end{figure}

In Fig. \ref{fig:eophase} we show the generic phase diagram for any even or odd layered structure. These phase diagram plots will serve as a guide on how we can explore the parameter space to study the thermal Hall behavior. We note that band crossings are a necessary, but not a sufficient condition for TPTs. Thus, to verify the existence of TPTs we explicitly compute the Chern numbers for each band in the gapped state for the required parameter set. If the Chern numbers rearranged themselves or changed values when the system became gapless under a parameter change, then we identified this band crossing as a TPT. While for the bi- and the trilayer each band crossing does in fact amount to a TPT, higher layer numbers do not always show this behavior. Therefore, carefully verifying that each crossing corresponds to a TPT is important. 

The edge states are the main source of novel phenomena in our multilayered system. Thus, determining the total number of edge states present within each phase is important to accurately characterize the physical response of each TPT. We do this by taking the sum of the winding numbers $\nu_n$, which are already partial sums of the Chern numbers. For our multilayers this can be expressed as
\begin{equation}
   \bar{\nu} =\sum\limits_{n=1}^{2L}\nu_n=\sum\limits_{n=1}^{2L} (2L - n)C_n,
\end{equation}
which are reported in the fourth column of Table \ref{tab:ChernTab} for the trilayer. 

To illustrate this concept, we provide an example of the determination of $\bar{\nu}$ for phase 1 of the trilayer. First, to determine the total number and chirality of the edge states in this phase, we calculate the winding numbers using the Chern numbers from Table \ref{tab:ChernTab}, given as $C_1=-1$, $C_2=+3$, $C_3=-1$, $C_4=-3$, $C_5=+1$, $C_6=+1$. Therefore the winding numbers are,
\begin{equation}
    \begin{aligned}
    &\nu_1=C_1=-1 &&=-1\\
    &\nu_2=C_1+C_2=\nu_1+3 &&=+2\\
    &\nu_3=C_1+C_2+C_3=\nu_2-1 &&=+1\\
    &\nu_4=C_1+C_2+C_3+C_4=\nu_3-3 &&=-2\\
    &\nu_5=C_1+C_2+C_3+C_4+C_5=\nu_4+1 &&=-1\\
    &\nu_6=0.
    \end{aligned}
\end{equation}
These numbers represent the number and chirality of the edge states that lie between each consecutive bulk band, with the knowledge that for all systems the final winding number is always zero. Therefore, by summing these numbers together, we get an idea about the net contribution of all the edge states present in that particular topological phase. For phase 1, this summation gives $\bar{\nu}=-1+2+1-2-1=-1$, in agreement with the value reported in the table. This process is repeated for each phase as the Chern numbers are rearranged. A comparison of each phase's net number of edge states has been done to understand the discontinuous behaviors which may appear as a result of the TPT.

{\setlength{\tabcolsep}{0.6em}
\begin{table}[!b]
 \centering
\begin{tcolorbox}[tab2,tabularx={c||@{\hskip 23pt}c@{\hskip 23pt}|c|c}]
\textrm{Phase} & $\overline{C}$ (\textrm{Chern Numbers})     & $\nu_3$ & $\bar{\nu}$ = $\sum_n \nu_{n}$           \\\hline\hline
1 & [-1, +3, -1, -3, +1, +1] & \multirow{2}{*}{1}& -1\\
2 & [+3, -1, -1, +1, -3, +1] & & 7 \\ \hline
3 & [+3, -1, +1, -1, -3, +1]  & \multirow{4}{*}{3} & 9\\
4 & [+3, -1, +1, -3, -1, +1] & & 7\\
5 & [+3, +1, -1, -3, +1, -1] & & 11\\
6 & [+1, +3, -1, -3, +1, -1] & & 9\\ \hline
\end{tcolorbox}
\caption{Chern numbers, the net number of edge states, and the number of edge states in the large gap for each phase labeled as they appear in Fig.~\ref{fig:TPT}b.}
    \label{tab:ChernTab}
\end{table}}

\begin{figure*}[htbp]
      \centering
      \includegraphics[width=17.8cm]{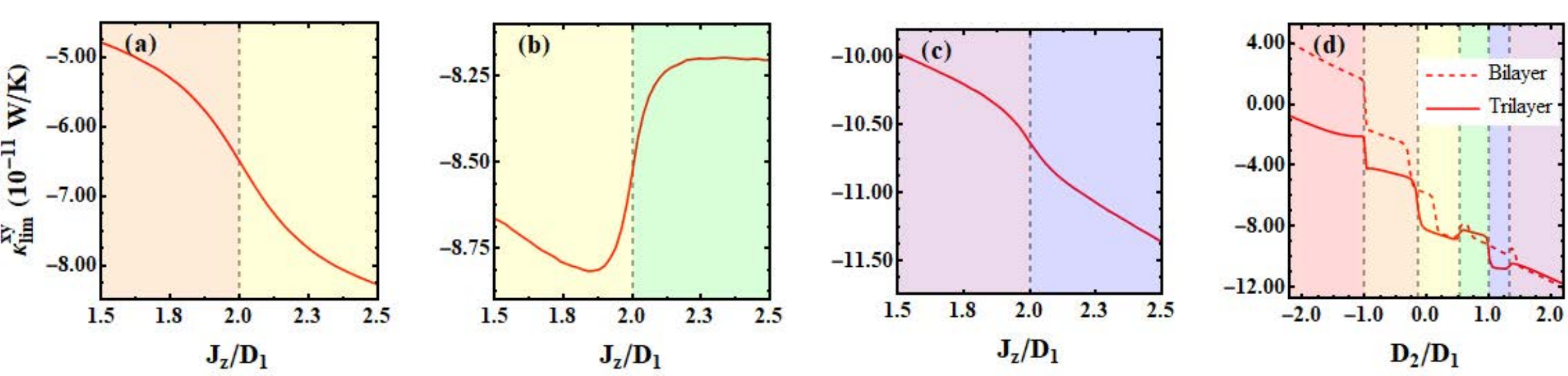}
      \caption{Thermal Hall conductance variation with Dzyaloshinskii-Moriya interaction and interlayer coupling.~\textbf{a} $\kappa_{lim}^{xy}$ as a function of $J_z/D_1$ with $D_2/D_1=-0.1481$, such that the transition occurs between phases 2 and 3 at $J_z/D_1 = 2$.~\textbf{b} $D_2/D_1=0.5322$ such that the transition occurs between phases 3 and 4.~\textbf{c} $D_2/D_1=1.3227$ such that the transition occurs between phases 6 and 5.~\textbf{d} $\kappa_{lim}^{xy}$ as a function of $D_2/D_1$ for the trilayer (solid red) and bilayer (dashed red). Both graphs are divided by the number of layers to normalize their contributions. The parameters are $S=\frac{1}{2}$ and $J=2J_z=4D_1$. Vertical dashed lines indicate the TPT points $\delta_n$ that seperate each phase. The phases 1-6 as shown in Fig. (\ref{fig:TPT} ) are ordered left to right.}
      \label{fig:TPTklim}
\end{figure*}

  \section{\label{sec:secIV}Thermal Hall Effect}
{\flushleft T}hermal Hall conductance is a useful response function to accurately characterize the topological nature of 2D magnonic materials~\cite{OwerreMono,MookHall,MurakamiPRB} and it given by\begin{equation}\label{eq:kappa}
      \kappa^{xy}=-\frac{k_B^2T}{(2\pi)^2\hbar }\sum_{n}\int_{BZ}c_2(\rho)\Omega^{xy}_n(\textbf{k})dk_xdk_y,
  \end{equation} 
with $c_2(\rho)=(1+\rho)\left(\ln\frac{1+\rho}{\rho}\right)^2-(\ln\rho)^2-2Li_2(-\rho)$, where $k_{B}$ is the Boltzmann constant, $\hbar$ is the Planck's constant, $T$ is the temperature, $\rho$ is the Bose-Einstein distribution, and $Li_2(\rho)$ is the polylogarithm function. We notice that the magnitude of the conductance is governed by both the weight function $c_2(\rho)$, where $\rho$ is the Bose-Einstein distribution, and the Berry curvature as calculated in Eq. \ref{eq:BC}. While the Berry curvature is primarily a function of the variables $J_z/D_1$ and $D_2/D_1$, $c_2(\rho)$ is a function of temperature $T$. Fig.~\ref{fig:TPT}a shows the parameter values of $J_z/D_1$ and $D_2/D_1$ over which we explore the topological properties of the multilayer. Thus, we can ask the question at what value of the temperature should the conductance be evaluated such that the non-trivial (if any) nature of the TPTs may be accurately captured? Because each band in our energy spectrum has a non-vanishing Chern number, and therefore a non-trivial Berry curvature, for every phase considered in our parameter space we would like to ensure that $c_2(\rho)$ captures their contribution. This can be achieved by taking $T$ as high as possible below the thermal disordering temperature of the multilayer. That is, we will take the high temperature limit as a figure of merit, with the caveat that within this approximation spin wave modes have not become completely thermally disordered to transition to a paramagnetic region. The high enough temperature ensures that every band has an equal occupancy as per the Bose-Einstein distribution. Therefore, in order to characterize the thermal Hall conductance response of our system we will use the high temperature limit of Eq.~\eqref{eq:kappa} given by (see deriviation in Appendix~\ref{sec:appendixB})~\cite{MookHall}

\begin{equation}\label{eq:hlim}
\kappa_{lim}^{xy}=\frac{k_B}{(2\pi)^2\hbar}\sum_n\int_{BZ}E_n(\textbf{k})\Omega^{xy}_n(\textbf{k})dk_xdk_y.
\end{equation} 

In practice, the high-temperature limit is bounded by the magnetic ordering temperature of the system. For the case of 2D CrI$_3$ the Curie temperature is 45K~\cite{Huang2017}. For this work we assume the magnetic ordering is mainly determined by $J$ and the tuning of $J_z$ and $D$ does little to effect this. 

In Fig.~\ref{fig:TPTklim} we show how the conductance varies as the system evolves through its topological phases. These TPTs can be explored by either tuning $J_z/D_1$ or $D_2/D_1$. First, we plot phase changes as a function of $J_z/D_1$ in Figs.~\ref{fig:TPTklim}a -- Fig.~\ref{fig:TPTklim}c. Each TPT is associated with a jump in the conductance, reminiscent of the Quantum Anomalous Hall effect present in electronic systems \cite{QAHE}. The relative increase or decrease in magnitude of the conductance due to these jumps can be explained by considering the number of edge states available on either side of the transition. Generally, more edge states yield a higher magnitude of the conductance, while fewer edge states result in a lower contribution to the magnitude. This is particularly observable in Fig.~\ref{fig:TPTklim}b, where the conductance shows a sharp decrease in magnitude. This can be explained by the difference in $\bar{\nu}$ between phases three and four, as shown in Table \ref{tab:ChernTab}. On the left side of the transition, phase 3 hosts nine different edge states, while on the right side of the transition phase 4 hosts seven, thus a difference of two. This decrease in available edge states coincides with the decrease in magnitude of the conductance, as fewer edge states are available to transport thermal energy. Furthermore intuitively, we can conclude that the conductance is sensitive to both the Berry curvature, from which $\bar{\nu}$ is derived by way of the Chern numbers, as well as the energy spectrum simultaneously. Thus, the exact value of the jump will depend on the rearrangement of the energy spectrum of the bands across the phase transition in addition to the change in the Chern numbers.

In Fig.~\ref{fig:TPTklim}d we plot the conductance as the system passes through a multitude of TPTs by varying $D_2/D_1$ for the bilayer and the trilayer.
The general trend is that the magnitude of the conductance increases as $D_2$ increases. 
For both layers the jumps can be characterized by $\bar{\nu}$, as done before. The relative increase or decrease in magnitude of the conductance at each TPT coincides directly with the relative increase or decrease of the number of edge states within each phase. To compare the results of the bilayer to the trilayer, we divided the conductance of each by the number of layers present in the system to determine the per layer contribution to the conductance. The number of jumps for the bilayer is different compared to the trilayer. We can attribute this fact to the differences in the topological phase diagram of the two systems. As shown in Fig.~\ref{fig:eophase}, the bilayer displays seven TPTs while the trilayer has only five. This is a consequence of the symmetries imposed upon the system by the choice of the stacking arrangement, explained below. In particular, we observe that the bilayer shows no TPT at the isotropic $\delta = 1$ point, but the trilayer does. By exploring this particular value of $\delta$ for different layering numbers $L$, we were able to determine the general nature of the TPT in even and odd layered structures with regards to the thermal conductance.

  \begin{figure*}[hbtp]
      \centering
      \includegraphics[width=17.8cm]{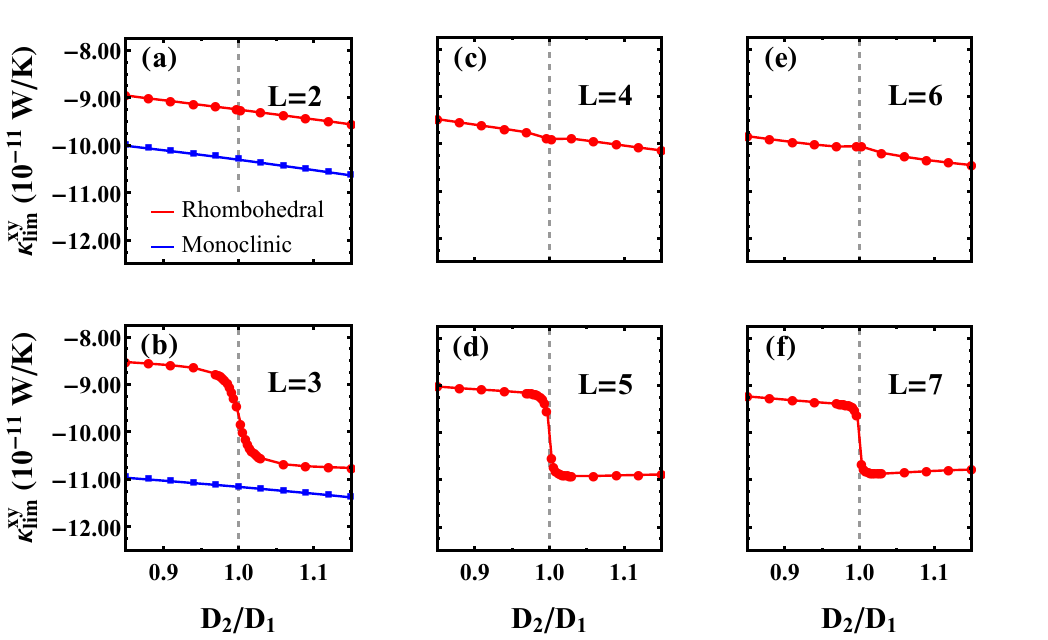}
      \caption{Thermal Hall conductance near $\delta=1$ for L=2 to L=7 multilayers.~\textbf{a}~-~\textbf{f}~Plots (a)~-~(f) are labeled by their value of L as shown in the plots. Even layered systems (top row) show no jump in the conductance, in contrast to odd layered structures (bottom row) which show a clear jump. Red circles indicate the rhombohedral stacking variation, while blue squares show the  monoclinic dependence.}
      \label{fig:oddeven}
\end{figure*}

In Fig. \ref{fig:oddeven} we show the conductance response at $\delta = 1$ for a set of few-layer systems, ranging from the bi- to the hepta-layer. It is clear that the even numbered layers show no jump in the conductance, while the odd numbered layers do. To highlight the sharpness of the jump across the TPT, for the odd layers, we chose a denser set of points near $\delta=1$. From Figs.~\ref{fig:oddeven}(b), ~\ref{fig:oddeven}(d), and ~\ref{fig:oddeven}(f) it is clear that the jump gets sharper as $L$ increases. Thus, within the limit of a few-odd layered systems, this effect is real and will survive. For even layers beyond $L=2$, band gap closings do occur for the same values of $\delta$ as the odd layers. However, these gap closings do not correspond to TPTs. 

The generalization of these results from the bi- and tri-layer to any layer can be shown by inspecting the analytically solved eigenvalues at $\pm K$. For $L$ layers, the Hamiltonian at $\pm K$ can be reduced to $L+1$ independent subspaces containing two $1 \times 1$ subspaces and $L-1$ $2 \times 2$ subspaces, which can be solved for their eigenvalues. Since the layers within each multilayer are structured such that their DMI strength alternates between the values of $D_1$ and $D_2$, the $2 \times 2$ subspaces will also repeat according to this pattern. Thus, beyond $L=3$ no unique subspaces occur, and subsequently no unique eigenvalues will be found. Therefore, the only distinguishable feature between multilayers will be the solution of the $(L+1)^{th}$ subspace, a $1 \times 1$ subspace which depends on the DMI value of the $L^{th}$ layer, $D_1$ (odd $L$) or $D_2$ (even $L$). By this reasoning we can categorize every multilayer by its even- or odd- ness. This generalization is shown in more detail in Appendix \ref{sec:appendixA}.

\begin{figure}[h]
     \centering
     \includegraphics[width=8.6cm]{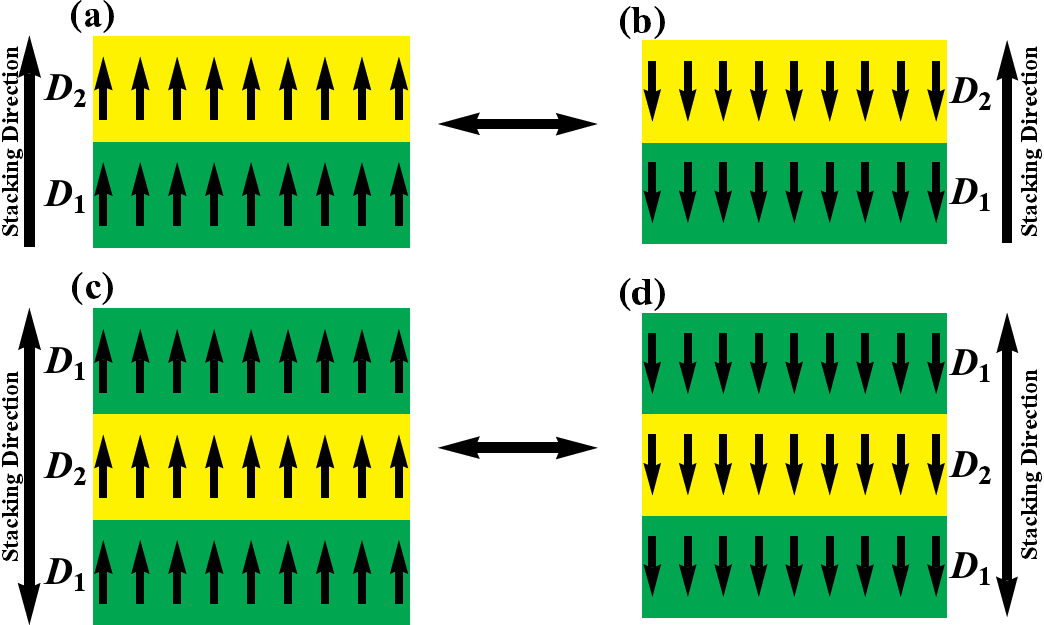}
     \caption{Bilayer (a) and its time-reversed (TR) partner (b), along with the Trilayer (c) and its TR partner (d). Each layer is labeled and colored by its Dzyaloshinskii-Moriya interaction term. The arrows within each layer denote the orientation of the spins. The time-reversal operation $\mathcal{T}$ flips the direction of the spin to produce a TR heterostructure. The stacking direction is denoted by the large vertical black arrow to the left or right of the heterostructure.}
     \label{fig:BiTriTRS}
 \end{figure}

The even-odd effect displayed in Figs.~\ref{fig:TPTklim}d and~\ref{fig:oddeven} is a result of the intrinsic spin orientation and the stacking direction which is imposed upon the structure by the choice of the stacking arrangement. The cartoon picture of bi- and tri-layer arrangement shown in Fig.~\ref{fig:BiTriTRS} demonstrates this principle. In the case of the odd-layered configuration the presence of a mirror symmetry imposed by the stacking arrangement works to preserve the invariance of the system under a time-reversal (TR) operation, which flips the spin orientations, as well as exchanging $+K$ and $-K$ in the Brillouin zone. This ensures that any gap closings must happen at $+K$ and $-K$ simultaneously. In contrast, even-layered configurations do not display this mirror symmetry, and therefore their solutions will not necessarily be TR-invariant. Therefore the even-layered configurations will host a higher number of topological band crossings as the crossings at $+K$ and $-K$ must be considered separately.

Finally, we note that recent Raman results suggest that room temperature mechanically exfoliated few-layer samples of CrI$_3$ retain their monoclinic structure even beyond the rhombohderal structural transition associated with the bulk material~\cite{Ubrig_2019}. In rhombohedral stacking each layer is associated with a shift of $a$ (the unit cell length) in the y-direction. In monoclinic stacking, the layers are shifted by $a/3$ in the x-direction. In Fig.~\ref{fig:oddeven}a and Fig.~\ref{fig:oddeven}b we show the results of the $\kappa^{xy}_{lim}$ for the monoclinic bi- and trilayer. In comparison to the rhombohedral case, neither multilayer shows a jump at $D_1/D_2=1$. This means that the odd-layered rhombohedral multilayers are the only configurations which show a jump at this parameter value. This indicates that the jump behavior is not universal.
 
 \section{\label{sec:secVI}Conclusion}In summary, we have studied the topological response of a multilayer configuration of hexagonal lattices stacked on top of each other in an ABC arrangement. Our calculation encompasses two different multilayer scenarios as characterized by their DMI interactions, only. While in general there may also be a pseudo-dipolar interaction term in the Hamiltonian, this term is not important for the class of systems considered here, and is therefore left for future study. We can have an all TMI system or another in which there is a combination of TMI-FM-TMI layers. For each of these setup, the observed topological phase transitions (manifested as jumps in the thermal Hall behavior) can be tuned by changing either the interlayer exchange interaction or DMI parameters. In an all TMI configuration the trilayer displays a jump in its thermal Hall conductance, while the bilayer does not. This even-odd jump response holds true beyond the bi- and trilayer. Thus, we propose a topological asymmetry layer experiment (TALE). By performing TALE one can decide whether an an asymmetric (even layered) or symmetric (odd layered) has been fabricated during the layering process. Such an experiment could potentially offer device fabrication physicists an additional means to characterize few-layered topological multilayer systems, besides the standard available methods~\cite{Mak2019}. We observe several topological phase transitions which are experimentally feasible since $J_z$ could be tuned ex-situ through various methods~\cite{Li2019,Song2019} allowing a continuous measurement through the TPT. The possibility to observe topological proximity effect and the presence of jumps distinguishing odd and even layers makes few-layered bosonic topological magnon systems an exciting playground to verify and apply fundamental concepts.
 
 Authors' note: During the writing of this article the authors become aware of a similar work wherein multilayers of dissimilar DMI were shown to host novel topological states in the form of chiral hinge magnons \cite{mook2020chiral}.

\begin{acknowledgments}
{\flushleft S}.~H and D.M would like to acknowledge funding from the NSF CAREER grant (ECCS, Award No.1846829) for support of this work. T.~D. acknowledges funding support from Sun Yat-Sen University Grant Nos. OEMT-2017-KF-06 and OEMT-2019-KF-04. S. T. acknowledges support from ARO Grant No. W911NF-16-1-0182 T.~D. thanks Jun Li and D.~X.~Yao for several helpful discussions.
\end{acknowledgments}

\FloatBarrier  
\appendix
\section{Topological Characterization}\label{sec:appendixA}
\begin{figure}[hbtp]
     \centering
     \includegraphics[width=8.6cm]{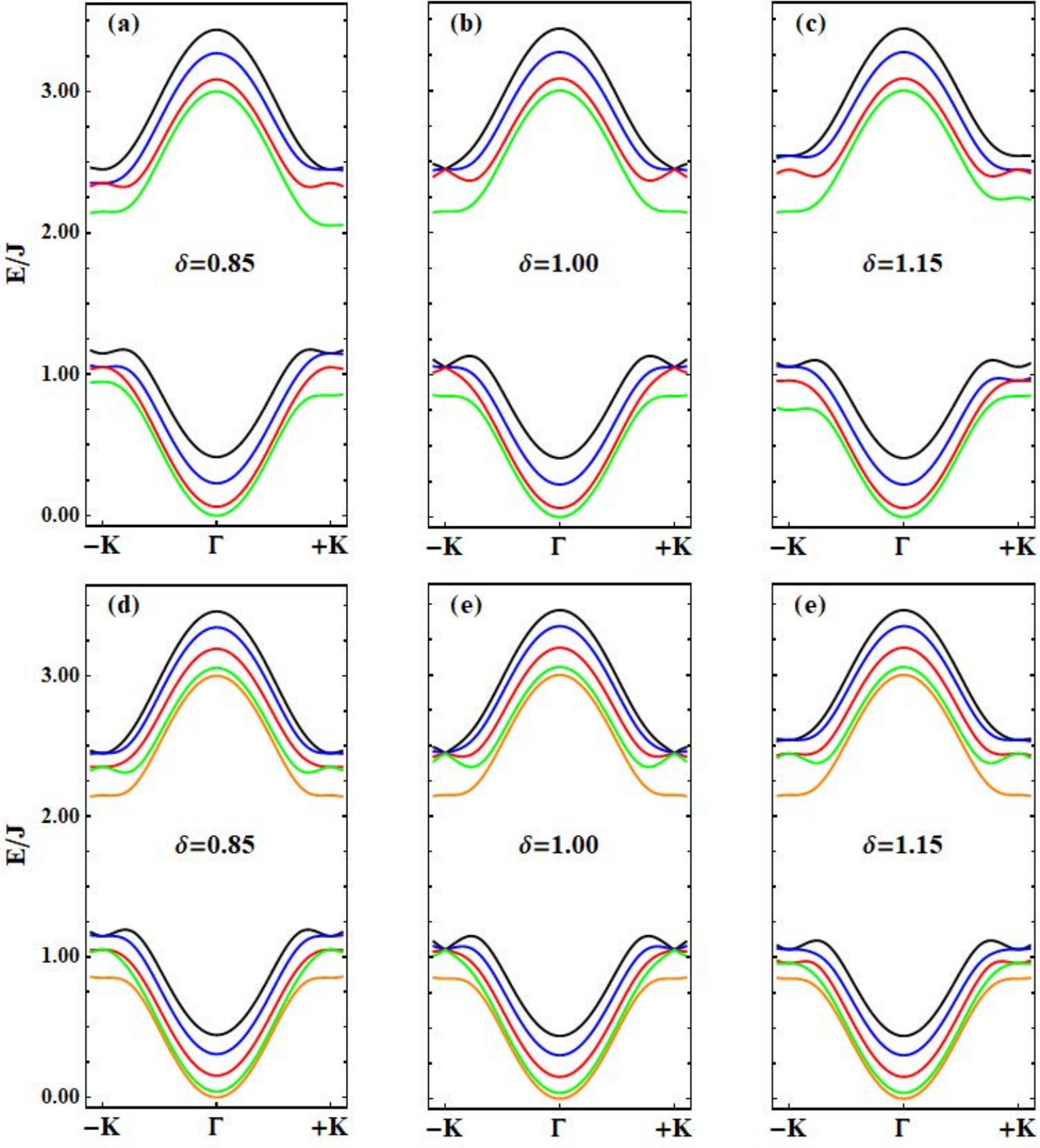}
     \caption{Eigenvalues of the $L=4$ (a)-(c) and the $L=5$ (d)-(f) structure along $k_y=0$ for values of $\delta$ around one (isotropic point). The y-axis represents energy in units of $J$.}
     \label{fig:L4and5}
 \end{figure}

Determination of TPTs is done by analyzing band crossings at the high-symmetry points $\pm K = \left(\pm\frac{4\pi}{3\sqrt{3}},0\right)$ in the Brillouin zone. By using the facts that $f(\pm K)=0$ and $m(\pm K) = \mp 3\sqrt{3}$, we can simplify the Hamiltonian and determine the eigenvalues analytically. In this case, Eq. \ref{eq:AMat} becomes,
\begin{equation}
\label{eq:AMatK}
     \mathcal{A}_{\tau}(\pm K)=\begin{pmatrix}\Theta^\tau_A\mp 3\sqrt{3}D_\tau S  & 0\\
    0 &\Theta^\tau_B \pm 3\sqrt{3}D_\tau S)
    \end{pmatrix},
 \end{equation}
 and Eq. \ref{eq:BMat} remains unchanged. The general form of the Hamiltonian evaluated at $\textbf{k}=\pm K$ can be written as, \begin{equation}
     \mathcal{H}_L(\pm K)= \begin{pmatrix} h_1^{(L)}& & & & \\
     & \ddots & & & \\
     & & h_\eta^{(L)} & &\\
      & & & \ddots &  \\
      & & & & h_{L+1}^{(L)}
     \end{pmatrix},
 \end{equation}
 resulting in $L+1$ subspaces. The Hamiltonian can be reduced to two $1\times1$ ($h_1^{(L)}$ and $h_{L+1}^{(L)}$ in the above equation) and $L-1$ $2\times2$ subspaces. Due to the alternating nature of our multilayers, the subspaces will likewise alternate resulting in the following general forms, \begin{equation}
     \begin{aligned}
     &h_1^{(L)}=3JS\mp 3\sqrt{3}D_1S, \\
     &h_{2l}^{(L)}=\begin{pmatrix}3JS+J_zS\pm3\sqrt{3}D_1S & -J_zS \\
     -J_zS & 3JS+J_zS\mp3\sqrt{3}D_2S
     \end{pmatrix},\\
     &h_{2l+1}^{(L)}=\begin{pmatrix}3JS+J_zS\pm3\sqrt{3}D_2S & -J_zS \\
     -J_zS & 3JS+J_zS\mp3\sqrt{3}D_1S
     \end{pmatrix},
     \\
     &h_{L+1}^{(L)}=3JS\pm3\sqrt{3}D_LS,
     \end{aligned}
 \end{equation}
 where the $2\times2$ subspaces are determined by the value of $\eta$ being even or odd. Eigenvalues for the bi- and trilayer systems are reported in Tables \ref{tab:EigL2} and \ref{tab:EigL3}. Due to the repetition of the subspaces, no eigenvalues at $\pm K$ beyond the $L=3$ system are found which are unique. Therefore, the main difference between even and odd layered multilayers is determined by the value of $D_L$, ($D_1$ for odd $L$ and $D_2$ for even $L$), which decides the eigenvalue for the $h_{L+1}^{(L)}$ subspace. In Fig.~\ref{fig:L4and5} we report the eigenvalues for the 4 and 5 layer configurations near the isotropic point $D_2/D_1=1$. 
 \begin{table}
     \centering
     \begin{tabular}{|c||c|c|}
     \hline
          &+K & -K  \\
          \hline \hline
         $E_1$ & $3JS -3\sqrt{3}D_1S$  &$3JS +3\sqrt{3}D_1S$ \\
         $E_2$ & \scalebox{0.89}{$J_zS +3JS - \frac{3\sqrt{3}}{2}D_1S(1-\delta) + S\Delta$} & \scalebox{0.89}{$J_zS +3JS + \frac{3\sqrt{3}}{2}D_1S(1-\delta) + S\Delta$}\\
         $E_3$ & \scalebox{0.89}{$J_zS +3JS - \frac{3\sqrt{3}}{2}D_1S(1-\delta) - S\Delta$} & \scalebox{0.89}{$J_zS +3JS + \frac{3\sqrt{3}}{2}D_1S(1-\delta)- S\Delta$} \\
         $E_4$ &  $3JS + 3\sqrt{3}D_2S$ &  $3JS - 3\sqrt{3}D_2S$\\
         \hline
     \end{tabular}
     \caption{Eigenvalues of the $L=2$ Hamiltonian evaluated at $\pm$K.}
     \label{tab:EigL2}
 \end{table}
 \begin{table}
     \centering
     \begin{tabular}{|c||c|c|}
     \hline
          &+K & -K  \\
          \hline \hline
         $E_1$ & $3JS -3\sqrt{3}D_1S$  &$3JS +3\sqrt{3}D_1S$ \\
         $E_2$ & \scalebox{0.89}{$J_zS +3JS - \frac{3\sqrt{3}}{2}D_1S(1-\delta) + S\Delta$} & \scalebox{0.89}{$J_zS +3JS + \frac{3\sqrt{3}}{2}D_1S(1-\delta) + S\Delta$}\\
         $E_3$ & \scalebox{0.89}{$J_zS +3JS - \frac{3\sqrt{3}}{2}D_1S(1-\delta) - S\Delta$} & \scalebox{0.89}{$J_zS +3JS + \frac{3\sqrt{3}}{2}D_1S(1-\delta)- S\Delta$} \\
         $E_4$ & \scalebox{0.89}{$J_zS +3JS + \frac{3\sqrt{3}}{2}D_1S(1-\delta) + S\Delta$} & \scalebox{0.89}{$J_zS +3JS - \frac{3\sqrt{3}}{2}D_1S(1-\delta) + S\Delta$}\\
         $E_5$ & \scalebox{0.89}{$J_zS +3JS + \frac{3\sqrt{3}}{2}D_1S(1-\delta) - S\Delta$} & \scalebox{0.89}{$J_zS +3JS - \frac{3\sqrt{3}}{2}D_1S(1-\delta)- S\Delta$} \\
          $E_6$ &  $3JS + 3\sqrt{3}D_1S$ &  $3JS - 3\sqrt{3}D_1S$\\
         \hline
     \end{tabular}
     \caption{Eigenvalues of the $L=3$ Hamiltonian evaluated at $\pm$K. Notice that the solutions of +K are the same as those of -K, but relabeled.}
     \label{tab:EigL3}
 \end{table} 

\section{Thermal Hall Weight Function}\label{sec:appendixB}
The transport properties of our heterostructure was characterized by the thermal Hall conductance~\cite{MurakamiPRB}\begin{equation}
      \kappa^{xy}=-\frac{k_B^2T}{(2\pi)^2\hbar }\sum_{n}\int_{BZ}c_2(\rho)\Omega^{xy}_n(\textbf{k})dk_xdk_y,
  \end{equation} 
with $c_2(\rho)=(1+\rho)\left(\ln\frac{1+\rho}{\rho}\right)^2-(\ln\rho)^2-2Li_2(-\rho)$, where $k_{B}$ is the Boltzmann constant, $\hbar$ is the Planck's constant, $T$ is the temperature, $n$ indexes the bands, $\rho$ is the Bose-Einstein distribution, and $Li_2(\rho)$ is the polylogarithm function. The weight function $c_2(\rho)$ favors low lying energy bands at low temperatures, while some of the topological phase boundaries in our system are defined by band crossings which occur at the higher end of our energy spectrum.  In order to characterize the topological phase diagram using the thermal conductance we must ensure the contribution of each band in the energy spectrum. Therefore, we opt to employ the high temperature limit of the conductance.
To find the high temperature limit $\kappa_{lim}^{xy}$ we write the above equation as \begin{equation}
      \kappa^{xy}_{lim}=\lim_{T\to\infty} \kappa^{xy}=\lim_{T\to\infty}\frac{-\frac{k_B^2}{(2\pi)^2\hbar }\sum_{n}\int_{BZ}c_2(\rho)\Omega^{xy}_n(\textbf{k})dk_xdk_y}{1/T}.
  \end{equation} Since $\lim_{T\to\infty}c_2(\rho)=\frac{\pi^2}{3}$ and $ C_n=\frac{1}{2\pi}\int_{BZ}\Omega^n_{xy}(\textbf{k})dk_xdk_y$, the numerator becomes \begin{equation}
  \begin{aligned}
     \lim_{T\to\infty} -\frac{k_B^2}{(2\pi)^2\hbar }\sum_{n}&\int_{BZ}c_2(\rho)\Omega^{xy}_n(\textbf{k})dk_xdk_y \\&=-\frac{k_B^2}{(2\pi)^2\hbar }\frac{\pi^2}{3}\sum_{n}2\pi C_n=0
     \end{aligned}
  \end{equation} where we have used the fact that $\sum_{n} C_n=0$. Additionally, since $\lim_{T\to\infty}(1/T)=0$, we can apply l'H\^opital's rule such that \begin{equation}
       \lim_{T\to\infty}\kappa^{xy}=\lim_{T\to\infty}\frac{k_B^2T^2}{(2\pi)^2\hbar }\sum_{n}\int_{BZ}\frac{\partial c_2(\rho)}{\partial T}\Omega^{xy}_n(\textbf{k})dk_xdk_y.
  \end{equation} To determine the partial derivative $\frac{\partial c_2(\rho)}{\partial T}$, remember that $c_2(\rho)=(1+\rho)(\ln\frac{1+\rho}{\rho})^2 -(\ln\rho)^2 -2Li_2(-\rho)$. If we define \begin{equation}
\beta = e^{\frac{E(\textbf{k})}{k_B T}};\hspace{25pt} \rho= 1/(\beta -1),
\end{equation} the derivative of the first term with respect to temperature becomes, \begin{equation}
    \begin{aligned}
   \dt{}\left(1+\rho\right)\left(\ln\frac{1+\rho}{\rho}\right)^2    =&\dbdt \big{[} \rho^2(\ln\beta)^2\\&-2(1+\rho)(\ln\beta)\beta^{-1}\big{]}.
    \end{aligned}
\end{equation} The second term gives us,
 \begin{equation}
    \begin{aligned}
  &\dt{(-(\ln\rho)^2)}=-\dbdt (2\rho(\ln\rho)).
    \end{aligned}
\end{equation} To determine the partial derivative of the third term we use the definition $Li_2(z)=\sum_{k=1}^{\infty}\frac{z^k}{k^2}$ to obtain, \begin{equation}
    \begin{aligned}
  &\dt{} \big{(}-2Li_2(-\rho)\big{)}=\dbdt \big{[}2\rho (\ln(1+\rho))\big{]}.
    \end{aligned}
\end{equation} Combining all the terms yields,
\begin{equation}
    \begin{aligned}
  &\dt{c_2(\rho)}= \dbdt\big{[}\rho^2(\ln\beta)^2-2(1+\rho)(\ln\beta)\beta^{-1}\\&\hspace{90pt}-2\rho(\ln\rho)+2\rho (\ln(1+\rho))\big{]}\\
  &=\dbdt\big{[}\rho^2(\ln\beta)^2-2(1+\rho)(\ln\beta)\beta^{-1}+2\rho (\ln\beta)\big{]}.
    \end{aligned}
\end{equation} 
Next, using the following limit expressions,
\begin{equation}
\begin{aligned}
&\lim_{T\to\infty} \beta =1;&&\lim_{T\to\infty} \rho = \infty;\\&\lim_{T\to\infty} \ln \beta = 0;&&\lim_{T\to\infty} \rho (\ln \beta) =1.
\end{aligned}
\end{equation} we have 
\begin{equation}
 \begin{aligned}
     \lim_{T\to\infty} T^2 \dt{c_2(\rho)}&=\lim_{T\to\infty}\Bigg{\{}T^2\dbdt\big{[}\rho^2(\ln\beta)^2\\&-2(1+\rho)(\ln\beta)\beta^{-1}+2\rho (\ln\beta)\big{]} \Bigg{\}}\\
    &=\frac{E(\textbf{k})}{k_B}.
     \end{aligned}
 \end{equation} Thus we have the final expression as\begin{equation}
     \kappa^{xy}_{lim}=\frac{k_B}{(2\pi)^2\hbar }\sum_n\int_{BZ}E_n(\textbf{k})\Omega^{xy}_n(\textbf{k})dk_xdk_y.
 \end{equation} 

  \FloatBarrier

\bibliography{apssamp}

\end{document}